# Blockchain-based Digital Twin for Supply Chain Management: State-of-the-Art Review and Future Research Directions

Jiongbin Liu[1], William Yeoh[2], Youyang Qu[1], and Longxiang Gao[1]

[1] School of Information Technology, Deakin University, 221 Burwood Highway, Burwood, VIC 3125 Australia

[2] Department of Information Systems and Business Analytics, Deakin University, 70 Elgar Road, Burwood, VIC 3125 Australia

Supply chain management (SCM) plays a vital role in the global economy, as evidenced by recent COVID-19 supply chain challenges. Traditional SCM faces security and efficiency issues, but they can be addressed by leveraging digital twins (DTs) and blockchain technology. The combination of blockchain and DTs can refine the concepts of both technologies and reform SCM to advance into Industry 4.0. In this paper, we provide a comprehensive literature review of the blockchain-based digital twin (DT) solutions to optimise the processes of data management, data storage, and data sharing in SCM. We also investigate the key benefits of the integration of blockchain and DTs and examine their potential implementation in various SCM areas, including smart manufacturing, intelligent maintenance, and blockchain-based DT shop floor, warehouse, and logistics. Finally, we put forward recommendations for future research directions.

## 1 INTRODUCTION

Since blockchain technology and DTs were first proposed in 2008 and 2012, respectively, they have been developing rapidly to revolutionise SCM through improving the security and efficiency of data processing, storage, and exchange. A DT is a digital representation of a physical asset, enabling engineers to perform a simulation to understand, evaluate, and predict behaviours for maintenance even before establishing the physical object [1]. However, the interaction between the physical and digital spaces will generate a great amount of data that need to be transmitted and stored in diverse infrastructures, such as clouds and fogs. These traditional data storage methods are considered to suffer from various security and privacy issues [1]. Therefore, the emergence of blockchain technology is ideal for improving the security and reliability of the DT data due to its decentralisation, transparency, and immutability nature. Additionally, the utilisation of a smart contract ensures the tamper-proof DT data and transactions within the supply chain system. According to Gartner, over two-thirds of companies implementing IoT would decide to deploy at least one DT in their production process [2].

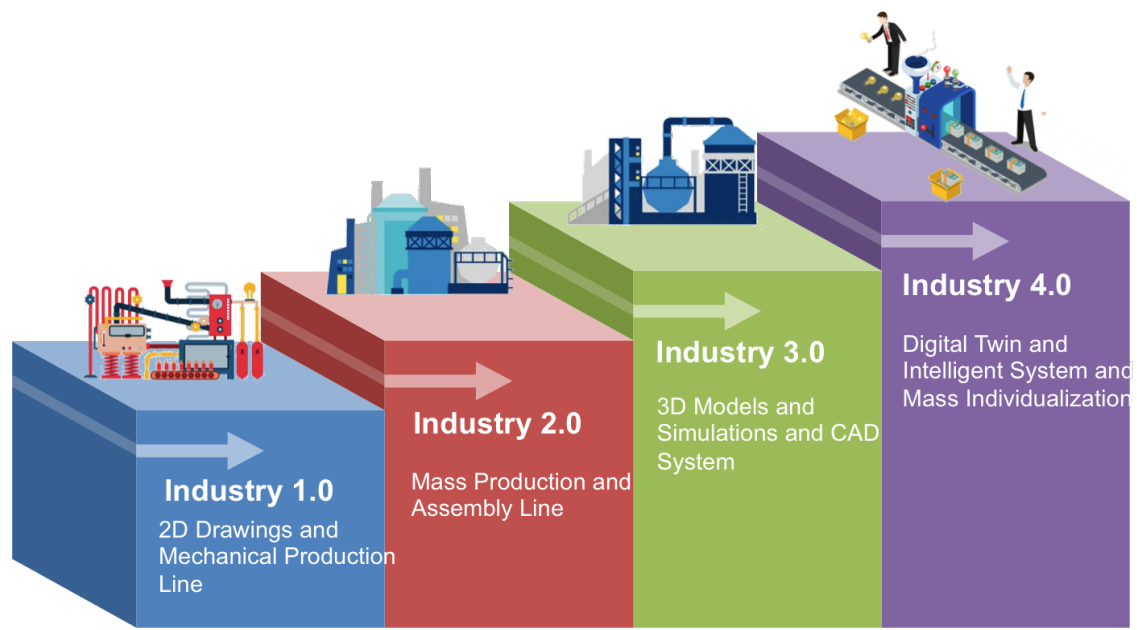

Fig. 1.  Evolution of the supply chain industry

The supply chain industry has evolved in four revolutions from Industry 1.0–4.0, as shown in Figure 1. In Industry 1.0 and 2.0, the design methods mainly utilised 2D drawings and physical prototypes, resulting in huge commissioning costs [3]. In Industry 3.0, the supply chain industry entered into the computer era with the development of computer-aided design (CAD) software that proposed the concept of digital prototypes and applied 3D model simulations to reduce the cost of failure [3]. In Industry 4.0, the supply chain is developed with mass individualisation and integration of the cyber-physical system and DTs [3]. As a consequence, the transactions among supply chain parties and the data transmission between the physical and digital spaces are becoming more challenging. Notably, the emergence of blockchain technology can benefit the DTs through improved security, traceability, transparency, and efficiency of DTs data processing. Therefore, this paper aims to conduct a comprehensive review of blockchain-based DT literature for SCM. Specifically, we identify four research questions that guide our literature survey:

1) RQ1: What are the definitions and characteristics of DTs and blockchains from the current literature?
2) RQ2: Which blockchain- and/or DT-related research papers for SCM were most instrumental in driving the development of the literature thus far?
3) RQ3: What are the key benefits and potential implementations of blockchain-based DT for various areas of SCM?
4) RQ4: What are the future research directions for integrating blockchain and DTs for

SCM?

To the best of our knowledge, there is limited literature survey on the key benefits and potential implementation of various aspects of SCM, leveraging DTs and blockchain technology. Existing surveys regarding blockchain and DTs did not focus on how the data processing benefited from integrating these two technologies and failed to investigate the potential implementation of blockchain and DTs in SCM.

In this paper, we provide a comprehensive literature review for the blockchain-based DT solutions to optimise the processes of data management, data storage, and data sharing. We also focus on the blockchain-based DT implementation for SCM, including the aspects of smart manufacturing, intelligent maintenance, and blockchain-based DT shop floors, warehouses, and logistics. The main contributions of this literature survey are as follows:

- This paper provides a comprehensive literature review for the blockchain-based DT solutions in addressing the data-related issues.
- This paper investigates the key benefits of integrating blockchain and DTs with SCM that can act as guidelines to facilitate further development.
- This paper analyses the implementation of blockchain and DT in various SCM areas to address the gaps in current literature.
- This paper provides recommendations for future research directions for implementing blockchain and DT in SCM.

Applying literature survey approach, we examined the most relevant and recent research papers. First, we defined three different strings of words to be searched on the databases: 'DTss' and 'blockchain' and 'supply chain' or 'supply chain management'. Since there is no published paper related to these three words altogether, we searched either 'digital twins' or 'blockchain' with 'supply chain' or 'supply chain management' or 'digital twins' and 'blockchain' only. A literature search on blockchain-based DT for SCM is conducted in the Web of Science database using the aforementioned search string. Additionally, we utilised the screening criteria to exclude duplicate papers and those not written in English. We include those published in a scientific journal, book chapter, or conference. The statistics of the collected literature with the three groups of keywords are shown in Figures 2-4.

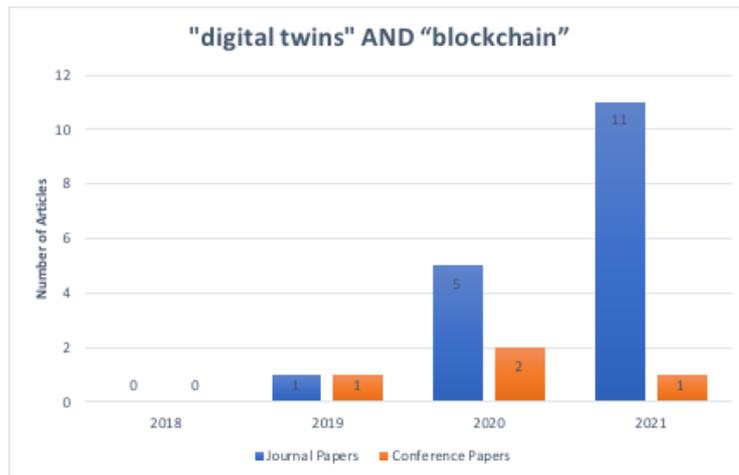

Fig. 2. Statistics of collected literature (digital twins and blockchain)

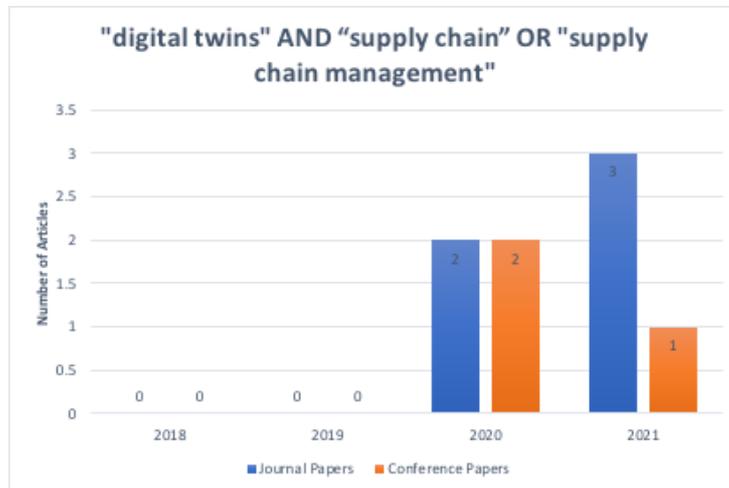

Fig. 3. Statistics of collected literature (digital twins and supply chain)

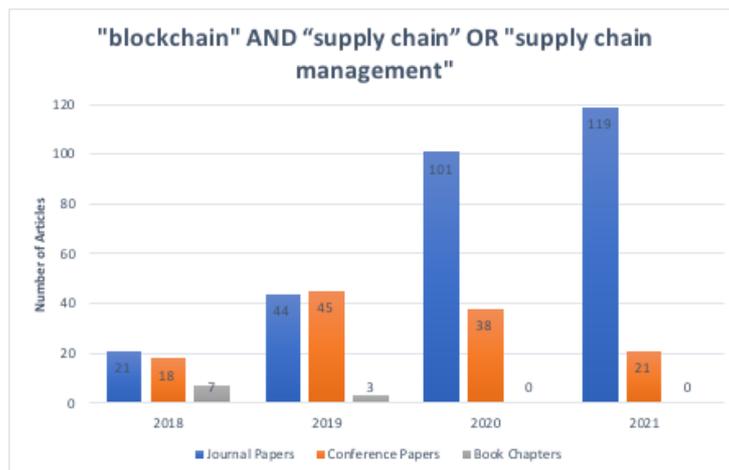

Fig. 4. Statistics of collected literature (blockchain and supply chain)

The remainder of this paper is organised as follows. Section 2 provides the background regarding DTs and blockchain, and presents the integration of DTs and blockchain to address data-related issues. Section 3 investigates the key benefits and the potential implementation of DTs and blockchain in various SCM areas. Section 4 outlines and discusses the future research directions. Section 5 concludes the survey.

## 2 BACKGROUND
### 2.1 Digital Twins

Grieves first introduced the concept of DTs in 2003 for the management of the whole product lifecycle in the virtual space [4]. There are various definitions for DTs collected from different sources. In 2012, the definition of DTs was proposed by the National Aeronautics and Space Administration (NASA), which refers to an integrated multi-physics, multi-scale, and probabilistic simulation, mirroring the state of a physical object in real-time based on historical data, physical model data, and sensor data [5]. Pethuru Raj [6] defined DTs as an exact virtual representation or replica of any tangible physical system or process. According to L Hou et al. [7], the concept of DT is to create a digital replica of a physical asset and then synchronise data generated from the digital and physical objects to monitor, simulate and optimise the physical asset. A DT is an essential component of the concept of the Cyber-Physical System [8, 9] that can be considered as any physical object, process, system, and the like that perform an interaction with a digital space through a communication network [10, 11].

A DT is not only a simple simulation model [4, 12-14]. It is also an intelligent model that can monitor, operate and optimise the lifecycle of its physical twin and continuously interact and predict future possible defects, failures, or damages. Additionally, DT can maintain the physical asset preventively through simulating and testing its digital configurations [11]. As shown in Figure 5, the DT model is defined through several components, including sensors and actuators from the physical space. Then, it creates, communicates and integrates with the digital space for data analysis and continuous updates. Additionally, a DT can act as a software program to input a variety of data from real-world scenarios and output useful simulations and predictions to enable an improved decision-making process [6].

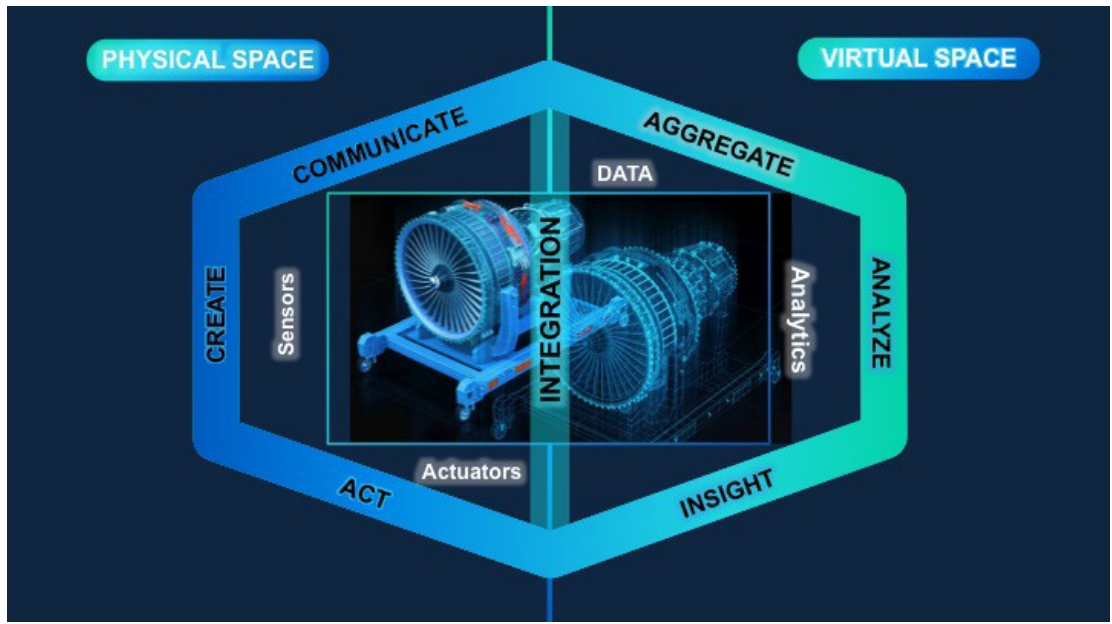

Fig. 5. Manufacturing process DT model

**2.2 Blockchain Technology**

Blockchain technology was first proposed as an underlying technology for cryptocurrencies, such as Bitcoinand Etherum [15, 16]. The development of blockchain has brought great changes to various domains, including finance, government, healthcare, and supply chains, thanks to its characteristics of security, decentralisation, immutability, transparency, traceability, and irreversibility. Blockchain can be defined as a distributed and decentralised append-only database by utilising cryptographic and computer networking techniques to ensure data storage and sharing security [17, 18], also acting as a platform for implementing digital transactions [19, 20].

A block's fundamental elements contain an index, timestamp, transaction data, and hashes of current and previous transactions. Then, two or more blocks are chained together, adopting asymmetric cryptography to enhance the security of transaction data [21]. Additionally, blockchain distributed ledgers allow each transaction to be verified by network nodes through a consensus mechanism before being added to the blockchain [22-24]. Along with the consensus mechanism, the smart contract is another essential component of blockchain, as described in the next sub-section.

**2.3 Blockchain-based Digital Twin**

As mentioned above, DTs enable engineers to analyse historical data in the past, track current performancesof the physical asset, and predict the state of the device in the future throughout the entire lifecycle. During each lifecycle phase of DT, the physical object sends various data to its DT counterpart. At the same time, blockchain technology can be utilised to securely document these data and provide real-time synchronisation and continuous updates due to blockchain technology's distributed, immutable, decentralisation, and trustless attributes. Furthermore, blockchain ensures provenance data to be tracked and traced on-chain, making it an effective solution to monitor the creation process of DTs [10, 25]. Blockchain enables DT data to be stored in a secure and tamper-proof ledger and ensures the efficient creation of virtual models [26]. Therefore, the development of blockchain technology can refine the DT process by ensuring transparent, efficient, and secure data management, data storage, and data sharing in industrial sectors, including SCM. A summary of blockchain for DT research is provided in Table 1.

Table 1. Summary of blockchain for DT research

| Papers | Focus Area | Problem Addressed | Main Approach | Key Contributions |
|---|---|---|---|---|
| [27] | Data management, data storage, and data sharing | Efficiency and security issues for managing DT data in a complex network | Proposed framework, case study | 1. Propose a data management method for DT of product based on blockchain.<br>2. Establish a peer-to-peer network and utilise a smart contract to enhance data sharing efficiency.<br>3. Integrate blockchain to improve data storage security. |
| [28] | Data management | Distrust between stakeholders and insufficient interactions between physical and digital spaces in manufacturing service collaboration | Conceptual, proposed mechanism | 1. Design a blockchain-based DT manufacturing service collaboration mechanism.<br>2. Explore DT and blockchain-enhanced manufacturing service management based on the industrial Internet platform. |
| [29] | Data management and data storage | Trustworthiness and security issues of data management and data storage | Proposed framework | 1. Apply data provenance and data twinning to enhance data trustworthiness.<br>2. Propose a blockchain-based framework to improve the security of data management and data storage. |
| [1] | Data storage | The lack of a common narrative about DT and | Proposed framework, case study | 1. Build a common narrative for DT by proposing a DT framework. |

| | | security and efficiency issues of the DT data storage | | 2. Design a blockchain solution to improve the efficiency of DT data storage. |
|---|---|---|---|---|
| [30] | Data storage and data sharing | The lack of a secure data sharing model for the interaction of DTs and lifecycle parties | Conceptual, prototype design | 1. Define a fine-grained access control for DT data sharing. 2. Propose a blockchain-based DT prototype for information management. |
| [31] | Data sharing | Integrity and confidentiality issues for secure DT data sharing | Theory-building, proposed framework, use case | 1. Explore the requirements for secure DT data sharing. 2. Propose a distributed ledger-based framework for secure DT data sharing. |

## 3  BLOCKCHAIN-BASED DIGITAL SUPPLY CHAIN TWIN (BC-DSCT)

Traditional SCMs are considered high risks due to the asymmetry of information communicated among variousparties in the ecosystem, the lack of transparency and visibility during production, the difficulty of control and traceability of actions in the distribution process, and the inaccuracy of information exchange within the supply chain parties. Several technologies have been implemented with the supply chain to mitigate these risks, including IoT devices, RFID technology, and ERP system. However, since they are centralised, these technologies make it hard to address certain problems, such as data tampering, privacy leakage, and trust issues. Therefore, the decentralisation feature and distributed ledgers of blockchain technology could improve the security and transparency in a trustless supply chain environment. Additionally, the integration of blockchain can build a resilient supply chain. For example, IBM has established blockchain-based supply chain solutions by using smart contracts to be automatically triggered when predefined business conditionsare met, providing near real-time visibility into operations [32].

The implementation of DT is another promising trend to improve the performance and efficiency of SCM.A digital supply chain twin is a digital replica of the physical supply chain consisting of lifecycle parties, including manufacturers, suppliers, distributors, and maintainers. The 3M company has created a DT to achieve standardisation and optimisation of their supply chain, and BMW is designing a complete DT of anentire factory that can be utilised to simulate 31 factories. The DT approach they applied will produce 30% more efficient planning processes [13]. Moreover, DT for SCM can improve the decision-making process through a range of advanced predictive and prescriptive analytics of models created for the end-to-end supply chain [33-35].

Increasing technologies could be leveraged together to achieve certain objectives in various industrial sectors, such as the integration of artificial intelligence (AI), machine learning, and blockchain technology to enhance IoT security. In this survey, we analyse the BC-DSCT concerning the key benefits of integration and the potential implementation for SCM in Industry 4.0, including smart manufacturing, intelligent maintenance, DT shop floor, and DT warehouses and distribution centres. These are described in detail in the following sub-sections.

## 3.1 Key Benefits of Integrating DT with Blockchain

*3.1.1 Enhanced Security and Fraud Prevention.* Secure data exchange and fraud prevention are two of the major objectives to improve the security of SCM. However, it is hard to realise in a centralised business environment because every point at which data is exchanged between supply chain parties presents an opportunity for it to be tampered with maliciously or inadvertently. Additionally, increasingly skilled and rampant counterfeits have brought an enormous loss and damage of reputation for the manufacturers. Thus, the integration of DT and blockchain with SCM could resolve the data security issue and combat the problem of counterfeits. On the one hand, DTs are implemented to record historical data and collect current data of all physical products and their transactions for analytics within the entire asset lifecycle. Further, the data stored on the blockchain can provide the integrity of DTs [30]. The real-time data generated from sensors deployed on the actual devices can help DT to perform a simulation to foresee security problems before they arise, which could also improve the data security of the DT supply chain. On the other hand, DT and blockchain are combined to ensure that counterfeit products are identified by providing product authenticity. For instance, there are a great number of replicas of luxury products with fake documents to prove their authenticity. Nevertheless, DTs can act as digital certificates stored on the blockchain to make them unable to be tampered with or stolen [6].

*3.1.2 Traceability of Digital Supply Chain Twin Data (DSCT).* In the past, it was extremely hard to track and trace which supply chain parties should take responsibility for the data or product issues, such as information leakage, malicious data tampering, and counterfeits. Progress has been made by introducing promising technologies, including RFID tracking and IoT devices, to address traceability issues. Meanwhile, the central and trust problems remain. Thus, the traceability of data exchanged among supply chain parties is further enhanced by integrating the provenance-enabled blockchain technology and the real-time data tracking between physical and digital spaces in DTs. The historical data is tracked through DTs to ensure the integrity and high accuracy of data in SCM. The current state of data is sent from multiple sensors deployed in the physical asset and received by the digital counterpart to obtain feedback for improvement. The future data is predicted for precaution of any possible malfunction and maintenance. These data are recorded on the tamper-proof blockchain and enable authorised supply chain stakeholders to track without any third party from multiple locations securely. Additionally, the cause of deviations between the physical and digital spaces can be identified by accessing DT data from the origin to the present to improve data security and authenticity [36].

*3.1.3 Transparency and Privacy-Preserving for DSCT Data Transaction.* In traditional SCM, an asymmetric information transfer is a primary issue that needs to be addressed as a priority, which would result in security risks, such as fraud and corruption, due to the lack of transparency. However, the supply chain parties are unwilling to disclose their information of transactions and DT data since improving transparency might cause a privacy leakage risk. Therefore, blockchain technology can be implemented to enforce transparency through smart contracts executed completely on blockchains, where the terms of the agreement are recorded on the distributed ledger as a chain code. Furthermore, the REST APIs can be used to develop blockchain applications and deploy blockchain networks that enable users to invoke a transaction and query or view the status and history of the transaction anonymously in their private profiles. At the same time, privacy can also be preserved through asymmetric encryption, that is, the utilisation of public keys and private keys to protect the transaction from DT data generated from product lifecycle based on the peer-to-peer supply chain network. Thus, the implementation of blockchain into the DSCT data transaction addresses the balance issue between enhanced transparency and privacy-preserving.

*3.1.4 Decentralisation and Immutability of DSCT Data Storage.* The prevalent technologies adopted in SCM, including RFID, the ERP system, and IoT, are operated in a centralised architecture, where data hacking and tampering risks exist. When DT data are transferred bidirectionally between many physical objects, consisting of physical assets, shop floors, warehouses and distribution centres in the supply chain industry, and their digital models are stored in centralised databases, similar issues might arise influence the security of DT data negatively. Therefore, blockchain is characterised as a promising technology that can enhance security through cryptographic hashing algorithms. Further, the DT data can be stored on distributed ledgers, which is decentralised, and anyone can verify the integrity of data recorded on the ledgers. However, the aforementioned different types of blockchain affect the degree of decentralisation. For example, a private blockchain is not fully decentralised like a public blockchain. Hence, the public blockchain is more secure than the private one concerning DT data storage. Additionally, blockchain ensures the immutability of DT data by remaining indelible and keeping the unalterable historical data of DTs. This benefits the DT audit process efficiently and cost-effectively, improves the level of trust significantly, and ensures the integrity of DT data [37].

*3.1.5 Blockchain-based Access Control on DSCT Data.* Traditional access control mechanisms are considered vulnerable to cyber security attacks. For instance, a number of security and privacy issues are found for the transaction in SCM due to the centralised approaches utilised to deal with access control mechanisms that rely on a third party. The approaches are constrained by availability and scalability. However, blockchain avoids unauthorised access that might cause malicious data manipulation, and only authorised supply chain parties to have access privileges to view, modify and simulate the DT data. Furthermore, blockchain-based access control strengthens security through secure socket layer certificates, which are deployed on each DT. Further, the certificate data is stored on blockchain [37]. Thus, integrating a tamper-proof blockchain with DT in the supply chain industry will enable fine-grained access control for only authorised supply chain parties permitted to take action on the DT data.

*3.1.6 Secure DSCT Data in a Trustless Blockchain System.* In traditional SCM, two separated profit groups, such as two downstream supply chain parties, have to share profits, so they would be unwilling to share information to gain a competitive advantage. The untrusted relationship between supply chain stakeholders has to rely on a third party during transactions. In some cases, over-reliance on intermediaries might cause a series of corruption problems, such as bribery or data manipulation. A more serious issue is that hackers tend to conduct cyber-attacks on these centralised organisations. Once they succeed, all clients' confidential information will be exposed, and the stolen data are difficult to recover since they are all stored in the systems of centralised organisations. Moreover, DT data need to be stored and utilised in a trustless environment even more, given that if the DT data is tampered with, the physical asset might be collapsed, and the shop floor might fall into disorder or even harm the entire supply chain operation. For this purpose, the trustless nature of blockchain that adopts distributed ledgers is a platform where supply chain parties can interact with each other in a purely peer-to-peer network without having to trust anyone in the blockchain system. Furthermore, the data generated from the physical machines, shop floor, warehouse, and distribution centre and their digital model data across the entire lifecycle can be stored, transferred, and simulated securely in a trustless blockchain network.

## 3.2 Potential Implementation in Industry 4.0

With a wide range of implementation-enabling technologies, including IoT, AI and cloud computing, and the rapid development of promising technologies, such as blockchain and 5G network in Industry 4.0, it is only a matter of time before the industry moves forward to a DT age [38]. In recent years, increasing organisations have developed DT solutions, integrating with various enabling technologies in various industries to optimise the performance of their physical assets, processes or systems. The Norwegian company, Visualiz, developed a DT start-up to establish a platform for the visualisation of physical assets in various industries. They integrated VR to create a collaborative space for users to monitor the status of assets online and detect the defects for maintenance [39]. Microsoft launched Azure DTs that allowed users to design digital models and knowledge graphs with the help of AI algorithms. Azure DTs enable users to model environments, such as a shop floor, warehouse, and distribution centre, by connecting assets like IoT devices and existing business systems [40]. Oracle develops a DT implementation based on IoT devices that operate on three pillars: virtual twin, predictive twin, and twin projections. This is done to address various issues. However, a virtual twin creates a virtual representation of a physical asset in the cloud, leveraging machine learning techniques. A predictive twin provides accurate predictions using a robust analytical model, and twin projections generate the predictions and insights to monitor the predicted state of the environment [40]. However, blockchain has the potential to become the most relevant and capable technology to address centricity, security, transparency and privacy issues in DTs [41, 43]. For example, in the car design process, blockchain is utilised for securely coordinating the activities or processes during the car design, including structural design, appearance design, stereoscopic measurement, and selecting the manufacturing method. The DTs of these processes perform interactions between the physical and digital spaces through smart contracts, which can achieve transparent and secure communications across the car design processes with their DTs [37].

Table 2. Mapping a catalogue for the potential implementation of blockchain and DTs in SCM

| Potential Implementation | Description | Digital Twin | Blockchain | References |
| --- | --- | --- | --- | --- |
| Smart Manufacturing | Apply smart machinery to monitor the production operation, and use data analytics to improve the | A virtual replica of the physical manufacturing machinery for collecting real-time DT data and performing simulations to | Blockchain stores and transacts data from machinery components and validates certificates and transactions in the manufacturing | [3], [44], [45], [1], [28], [46], [47], [48] |
| | manufacturing performance. | optimise the performance. | process. | |
| Intelligent Maintenance | Use data collected from machinery to perform predictive and preventive analysis of defects. | A digital replica of the physical objects for analysing and predicting the health status and performing maintenance in advance. | Blockchain manages lifecycle data and stores different types of data to improve the accuracy and reliability of predictive maintenance. | [49], [50], [51], [52] |
| Blockchain-based DT Shop Floor | A factory and a basic unit of manufacturing for manufacturers to operate the machines to complete tasks | The interaction between the physical and virtual shop floor enhances security and efficiency in the manufacturing process. | The decentralised nature of blockchain protects data by storing data in shop floor blockchain and service blockchain separately. | [53], [54], [55], [56], [57] |
| Blockchain-based DT Warehouse and Logistics | Finished goods or raw materials stored in a warehouse are transported inbound and outbound for production or sale. | DTs optimise warehouse operations and management, reduce the collection and distribution time and management costs, and provide insights during shipment and transportation. | Blockchain improves the security and efficiency of warehouse data management and enhances transparency in the logistical process by tracking all needed information on the blockchain network. | [58], [59], [60], [61], [62], [63], [64], [65] |

Concerning the implementation of DT and blockchain into the supply chain industry, we can define the DSCT as a real-time simulation model of a physical supply chain that can predict the behaviour and dynamics of SCM to make effective decisions. By integrating DT and blockchain with SCM, the authenticity of products is guaranteed, and the counterfeiting of goods is eradicated effectively so that supply chain parties can save money and increase revenue significantly. DTs usually act as digital certificates stored on blockchain to ensure the data embedded in the certificates cannot be copied, tampered with, or deleted by hackers. Furthermore, the digital certificate will keep updating within the entire product lifecycle since the product is designed and manufactured. The shipment data transferred from multiple IoT sensors deployed in the containers will be updated with timestamps during the transportation stage. Hence the customers can track and trace the authenticity of the product by verifying its DT data stored on the blockchain [6].

General Electric (GE) is one of the industry pioneers implementing DT to collect data from its manufacturing, maintenance, operations, and operating environments. It utilises this data to create a unique model of each of its assets, systems or processes. The objectives of GE to develop DT solutions are to increase reliability and availability, reduce risk, lower maintenance costs, improve production, and achieve faster time-to-value [66, 67]. Additionally, the 3M company draws up a future project regarding the development of DT for their supply chain to optimise, visualise, and control all their processes [68]. BMW also endeavoured to develop DTs of their factories to optimise their performance [13]. Therefore, the following sub-sections focus on combining DT and blockchain into various aspects of the supply chain, as summarised in Table 2, including smart manufacturing, intelligent maintenance, DT shop floor, and DT warehouse and logistics. This was done through analysing their current research development and potential practical implementation in Industry 4.0. Besides, different research approaches applied in existing papers for these focus areas are catalogued in Table 3.

Table 3. Cataloguing research approaches in various focus areas

| Focus Areas | Proposed Framework | Case Study | Conceptual | Designed Prototype |
|---|---|---|---|---|
| Smart Manufacturing | [3], [1], [46], [28] | [1], [47] | [3], [44], [45], [47], [48] | [46] |
| Intelligent Maintenance | [51], [52] | [51] | [49], [50] | N/A |
| Shop Floor | [54], [55], [53] | [54], [55], [56], [57] | [53], [56], [57] | N/A |
| Warehouse and Logistics | [61], [62], [63] | [58], [61], [63], [65] | [60], [62], [64], [65] | [58], [59], [61], [63] |

*3.2.1 Smart Manufacturing.* Smart manufacturing is a technology-driven approach that applies Internet-connected machinery to monitor the production process. The objective of smart manufacturing is to achieve automating operations and utilise data analytics to improve the performance of the manufacturing process. However, a DT acts as a virtual replica of the real-time status of the manufacturing machinery. A great number of multi-faceted sensors and actuators are embedded with these machineries across the entire physical manufacturing process. Data collected from the sensors are continuously transmitted to the DT application, which would instantly analyse the data to detect any defects [69]. In this case, blockchain can be utilised to store and transact records of these tested data sets for a particular component of the machinery. Then, the DT model could perform simulations and predictions for the component to improve performance and maintenance. Furthermore, blockchain provides append-only functionality for the validation of certificates and transactions in the manufacturing process [47].

Table 4. Summary of BC-DSCT research related to smart manufacturing

| Papers | Highlights | Main Technologies | Key Findings |
|---|---|---|---|
| [3] | DT-based smart manufacturing system design | Digital twin | A framework is proposed to review how DTs are integrated into and facilitate the smart manufacturing system design. |
| [44] | Integration of DT with the control of the Manufacturing Execution Systems | Digital twin | A DT environment in a laboratory is proposed to overcome the discrepancies between literature and implementations of DT. |
| [45] | Comparison and integration of DT and big data in smart manufacturing | Digital twin, big data | The concepts, similarities, differences, and complementarity of DT and big data in smart manufacturing are reviewed and discussed. |
| [1] | A blockchain solution for the secure and efficient management of DT data | Digital twin, blockchain | A spiral DT framework is presented for building a common narrative for DT, and a blockchain solution, twin chain, is proposed for optimising DT data management. A case study is conducted to deploy the twin chain in smart manufacturing. |
| [28] | DT- and blockchain-enhanced smart manufacturing service collaboration | Digital twin, blockchain | A DT- and blockchain-enhanced manufacturing service collaboration mechanism is proposed to address the trust issue between stakeholders for the process of cyber-physical integration in the Industrial Internet platform. |
| [46] | Combination of IoT with blockchain and the configuration of a data- and knowledge-driven DT manufacturing cell | Blockchain, IoT, digital twin | A manufacturing blockchain of architecture for the configuration of a data- and knowledge-driven DT manufacturing cell is proposed to ensure the security, traceability and decentralisation of intelligent manufacturing systems. A blockchain prototype is established to implement the proposed architecture. |
| [47] | Establishment of a blockchain-based DT additive manufacturing in the aircraft industry | Digital twin, blockchain | Phases of the metal additive manufacturing process are discussed to further a DT in the aircraft industry. A case study is designed to integrate blockchain with additive manufacturing in the aircraft industry. |
| [48] | Exploitation of blockchain for | Blockchain | Measurements for the implementation of blockchain applications in the |

| | |
|---|---|
| solving cybersecurity issues in the manufacturing system | manufacturing system are discussed to overcome the cybersecurity issues, and reference architecture of the blockchain-secured smart manufacturing system is proposed to standardise blockchain computing. |

Figure 6 illustrates the blockchain-based DT assembly process, which is a crucial and final stage of smart manufacturing. It ultimately determines the quality of a complex product. The DT-driven assembly process based on IoT achieves the in-depth integration of the physical and virtual spaces and realises the precise control of components, equipment and assembly processes through the implementation of intelligent software service platforms and tools. Furthermore, the DT-driven product assembly process is uniformly and efficiently controlled to realise the product assembly system's self-organisation, self-adaption, and dynamic response.

One of the most important applications of DT- and blockchain-based smart manufacturing in Industry 4.0 lies in the operation monitoring and controlling for complex products that can realise intelligent interconnection, especially high-end intelligent equipment. The real-time sensor data collected from the operation of machinery and stored on the blockchain is transferred to its DT model for simulation analysis to diagnose the machinery's health status and malfunction risk. If the operating conditions of the product change, the DT model will be virtually verified on the blockchain platform. If there is no problem, then the operating parameters of the physical product can be adjusted. However, the integration of DT into smart manufacturing has been applied widely in the aviation industry.

Table 4 discusses the main technologies applied and the key findings of BS-DSCT research related to smart manufacturing. The introduction of DT into smart manufacturing can improve quality management by continuously monitoring product data collected from the embedded IoT devices. It can perform better product design by detecting design flaws in advance and enhancing supply chain efficiency through a deeper and more decisive view on materials usage and lean and just-in-time manufacturing realised by DT [66]. Furthermore, the implementation of blockchain technology can also be utilised in the aircraft industry to create a DT for this complex manufacturing process. Additionally, blockchain is an effective technology to address the cybersecurity issues of the manufacturing sector, such as the traceability of operations [70], cyber-attacks to the digital thread [71], device spoofing and false authentication in data sharing [72], interoperability, among heterogeneous equipment [43], confidentiality and trust between participants [73], information vulnerability and reliability across systems [74], and the failure of key nodes in centralised platforms [75].

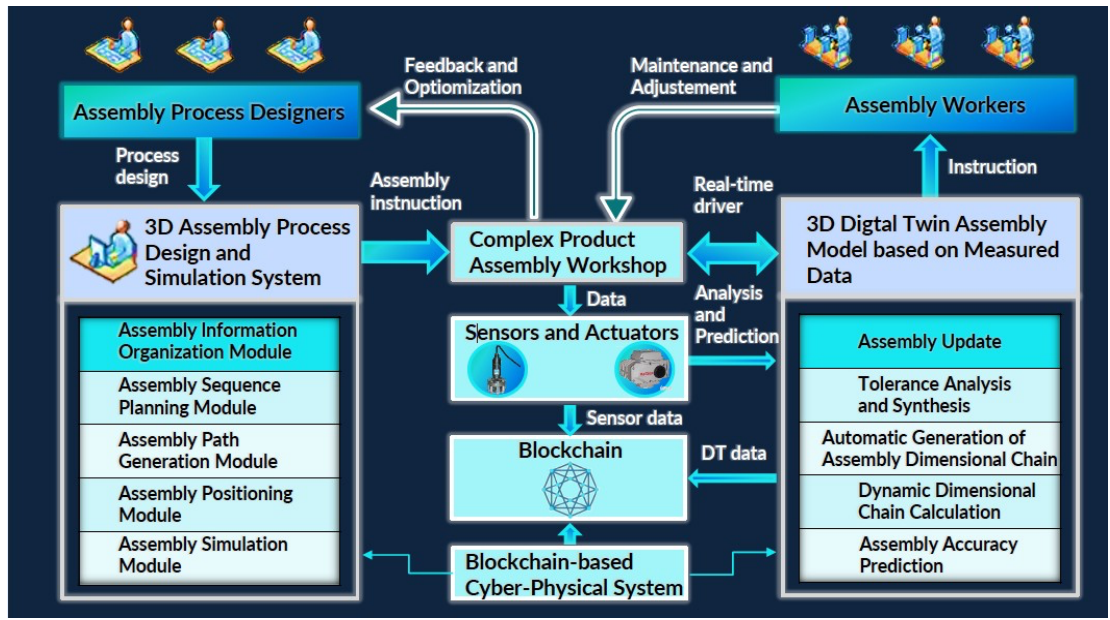

Fig. 6. Blockchain-based DT-driven complex product intelligent assembly

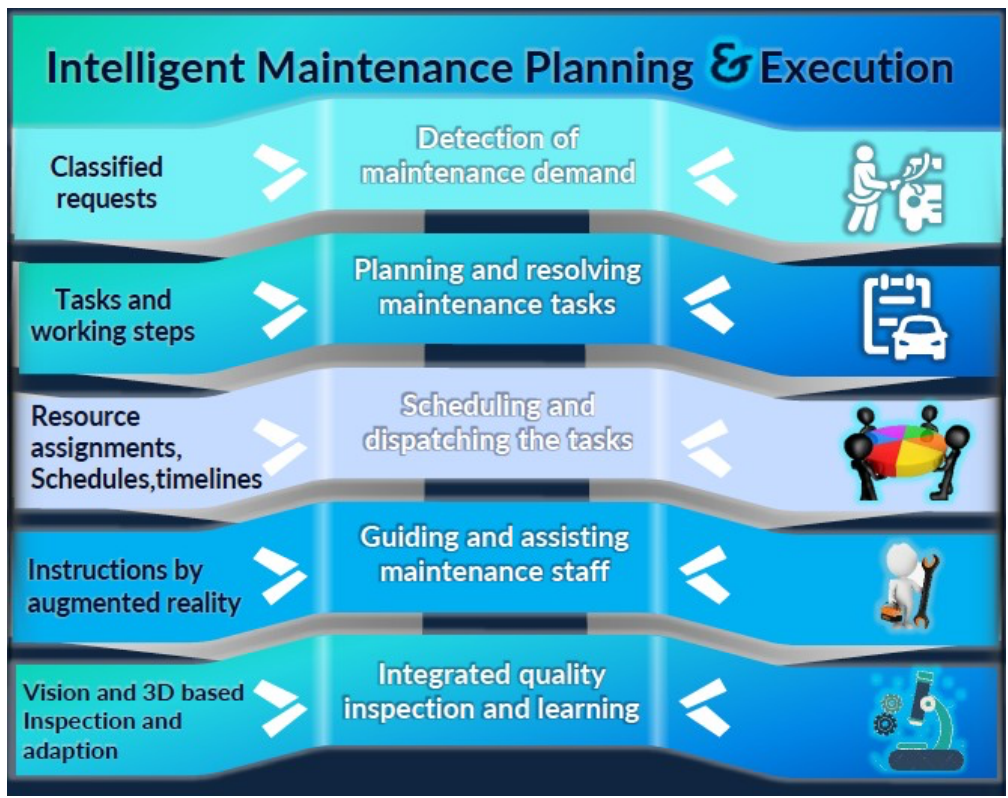

Fig. 7. Intelligent maintenance planning and execution

*3.2.2 Intelligent Maintenance.* Intelligent maintenance is defined as a system that utilises data collected from machinery to perform predictive and preventive analysis of potential defects or failures for the assets, systems, or processes. Figure 7 shows that the intelligent maintenance planning and execution process can be divided into five steps: detection, planning, scheduling, guiding, and quality inspection and learning. The first step is related to the detection of the maintenance demand and classification into specific maintenance cases, which contains the collection of production and sensor data. Based on the detected maintenance case, the second step will generate the work plan to evaluate the degree of urgency for scheduling the maintenance. Step three is selecting the best resource and finding an optimal schedule. Step four is related to the guidance and assistance provided to the maintenance personnel during the execution of the maintenance procedures. In the last step, quality inspection and learning will be conducted to correct the applied procedures and information regarding the maintenance, including unknown maintenance cases and materials or tools updates [76].

Table 5. Summary of BC-DSCT research related to intelligent maintenance

| Papers | Highlights | Main Technologies | Key Findings |
|---|---|---|---|
| [49] | Design of intelligent maintenance systems for the optimisation of maintenance operations | N/A | Intelligent prognostic and health management are utilised to identify effective maintenance strategies. Examples of system-level strategies are illustrated to present the effectiveness of the optimised maintenance operations. |
| [50] | DT applications for maintenance | Digital twin | Different maintenance strategies and the application of DTs in these strategies are discussed to fill the gaps in the current literature. |
| [51] | DT-driven predictive maintenance of enhancing reliability and the accuracy of the CNC machine tool | Digital twin | A DT model-based and DT data-driven hybrid approach is proposed to realise reliable predictive maintenance of the CNC machine tool. A case study regarding the lifecycle prediction of the cutting tool is conducted to demonstrate the feasibility and accuracy of the approach. |
| [52] | Combination of blockchain and DT for the intelligent maintenance of complex equipment | Blockchain, digital twin | An intelligent maintenance framework is proposed to utilise blockchain for the creation of DT and protection of data and offers blockchain and DT based solutions for the intelligent maintenance of complex equipment. |

Concerning DT for predictive maintenance, a digital replica of the underlying asset, process, or system can be utilised to analyse and predict the health status and perform maintenance in advance to ensure all operations run normally. Table 5 discusses the main technologies applied and key findings of BS-DSCT research related to intelligent maintenance.

The company Ansys devotes to developing Multiphysics engineering simulation software for product design, testing and operation. It provides DT services to wind power companies to avoid unplanned downtime and to achieve predictive maintenance, operation control, and optimisation [77]. Further, Ansys and GE have cooperated to promote DT applications to perform product lifecycle management, remote diagnosis, and intelligent maintenance to help customers analyse specific working conditions and predict faults through leveraging sensor data and simulation technology. They also combine the failure modes in the historical maintenance data of the engine with the three-dimensional structure model and performance model to establish a failure model for the fault diagnosis and prediction [78]. However, the integration of blockchain and DT with intelligent maintenance can realise condition motoring, fault diagnosis and life prediction of complex equipment. Besides, blockchain can be utilised to manage the lifecycle data and store different types of data, including component data, installation data, and operation process data for predictive maintenance [52]. Blockchain also applies cryptographic technology and distributed consensus protocols to enhance the security of DT data transmission, thereby improving the accuracy and reliability of predictive maintenance.

Table 6. Summary of BC-DSCT research related to blockchain-based DT shop floor

| Paper | Highlights | Main Technologies | Key Findings |
|---|---|---|---|
| [53] | A DT shop floor realising interaction and convergence between physical and digital spaces | Digital twin | A DT shop-floor paradigm is proposed to promote the development of smart manufacturing by building a conceptual model of a DT shop floor, including the interactions of a physical shop floor, virtual shop floor, shop floor service system, and shop floor DT data. |
| [54] | Shop-floor DT acts as a digital mapping model of the physical shop floor that needs to be built and applied properly. | Digital twin | An implementation framework is proposed for constructing and applying shop floor DT to model the objects and dimensions, monitor the real-time status, and predict the future status. A case study is conducted regarding the development of a DT-based visual monitoring and prediction system for the shop floor operating status. |
| [55] | Integration of DT with the complex product assembly shop floor for smart manufacturing management and | Digital twin, big data | A DT-based smart production management and control framework is proposed for the interactions among physical and DT assembly shop floor, as well as DT and big data-driven assembly shop floor service, storage, and management platforms in a satellite assembly shop floor. |

| | | | control |
|---|---|---|---|
| [56] | Combination of MTConnect and the Manufacturing Execution System for data collection in the shop floor DT | Digital twin, cloud computing | The development and implementation of a novel Manufacturing Execution System utilising mobile devices and cloud computing are proposed to integrate MTConnect for data collection. A case study is conducted to combine MES and MTConnect in a manufacturing facility. |
| [57] | Integration of blockchain with the maintenance in the shop floor using smart contracts | Blockchain | A case is discussed for the maintenance of manufacturing machines, and a blockchain solution is proposed to separate a blockchain from the shop floor blockchain and service blockchain for data protection and data security. |

*3.2.3 Blockchain-based Digital Twin Shop Floor.* A shop floor is the area of a factory, machine shop, and abasic unit of manufacturing, where manufacturers work on machines to complete the tasks. However, with the development of promising technologies in Industry 4.0, including IoT, big data, AI, cloud computing, and blockchain technology, the traditional shop floor has evolved to the virtual or digital generation based on the cyber-physical system. Thus, the interaction between the virtual and physical shop floors can significantly improve the security and efficiency in the manufacturing process. Tao and Zhang [53] first proposed a new shop floor paradigm, namely a digital twin shop floor (DTS), to achieve the convergence of shop floors in the physical and virtual spaces. The paradigm consists of four key components: physical shop floor(PS), virtual shop floor (VS), shop floor service system (SSS), and shop floor DT data (SDTD), as shown in Figure 8. Concerning their relationships, PS strictly follows the predefined orders from VS and SSS and organises production. Meanwhile, VS provides control orders for PS and optimises strategies for SSS. SSS integrates various computer-aided tools, models, and algorithms to form composite services for specific demands from PS and VS. SDTD fuses data of the previous three components and the existing methods formodelling, optimising and predicting, which can also eliminate the information isolated island for more comprehensive and consistent information. Table 6 discusses the main technologies applied and key findings of the BS-DSCT research related to the blockchain-based DT shop floor.

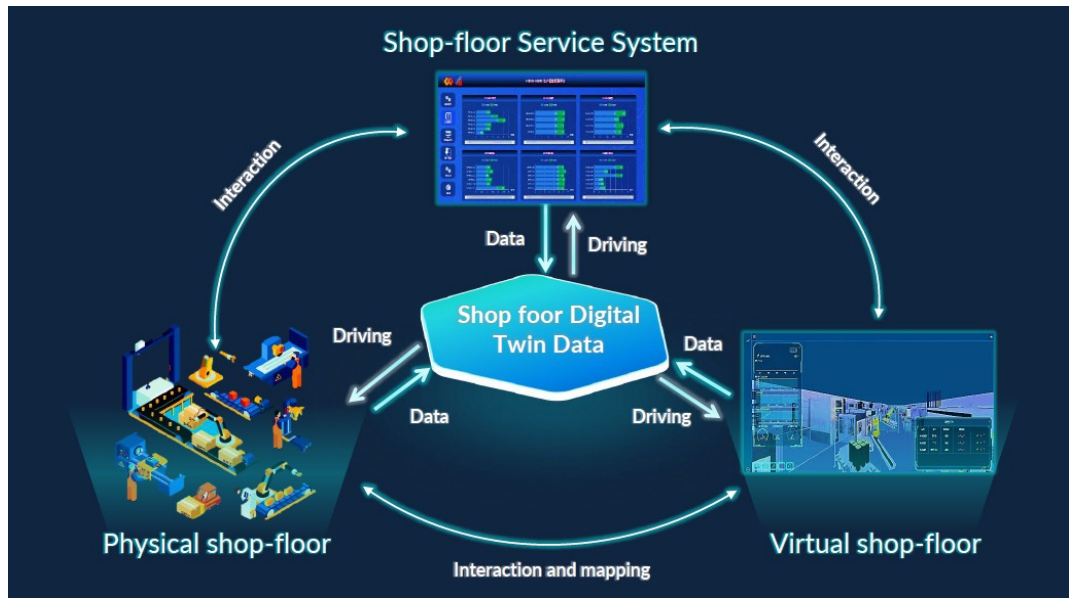

Fig. 8. Conceptual model of DTS

However, regardless of whether big data or cloud storage is used, they are centralised and might cause security issues, such as data tampering, malicious deletion, and data fraud. This could result in data inconsistency between the physical shop floor and the DTS or even cause disorder or stagnation of the shop floor. Therefore, blockchain technology has transformative power in the way of data protection, thanks to its decentralised architecture. Data can be stored in two types of blockchains, a shop floor blockchain and service blockchain, based on different levels of detail of the stored data to achieve data protection. The shop floor blockchain contains detailed information of the data generated from the physical shop floor and the DTS, including machine logs and maintenance reports. Meanwhile, the service blockchain is provided to different service providers and contains partial information, such as the maintenance reports without confidential data or internal data and tokenised artefacts like machine learning models. The advantage of the separation of the blockchain lies in the removal of unnecessary data stored in the service blockchain that could enhance data security, improve blockchain scalability, and reduce data storage costs. For instance, theservice blockchain is not affected by a great number of transactions imposed by storing the machine data, which is only stored in the shop floor blockchain [57].

Table 7. Summary of BC-DSCT research related to blockchain-based DT warehouse and logistics

| Papers | Focus Area | Highlights | Main Technologies | Key Findings |
|---|---|---|---|---|
| [58] | Warehouse | 3D packing and storage assignment in the warehouse can be optimised in the DT system | Digital twin | A DT-driven joint optimisation approach is proposed to aggregate real-time data for mapping the physical warehouse to the digital model. A prototype of the DT system is built with a case study to demonstrate it. |
| [59] | Warehouse | A warehouse management system with UAV based on DT for intelligent and controllable warehouse management | Digital twin, cloud, 5G | A DT-based warehouse management system is developed to upload the data and information of the warehouse to the cloud using 5G communication, which is then sent to the DT platform for the collection of real-time warehouse data and feedback for the operation of the UAV. |
| [60] | Warehouse | Integration of IoT-based smart warehouse with blockchain | IoT, blockchain | A hybrid model for an end-to-end data warehouse solution is proposed to implement IoT blockchain and smart contract, and the comparisons between centralised and decentralised approaches and the proposed model are discussed. |
| [61] | Logistics | A physical and cyber logistics system for a DT-based supply chain control | Digital twin | A cyber-physical logistics system is proposed with a multi-level architectural framework to provide technical functionalities on DT simulation for resilient supply chain control. |
| [62] | Logistics | A DT-based and simulation-based decision support tool for the in-house logistics system | Digital twin | A simulation-based decision support tool is proposed to develop the architecture of a DT for in-house logistics and simulation models for analysing logistics activities for both distribution and production facilities. |
| [63] | Logistics | Combine DT into the logistic production system to systematically | Digital twin, cloud computing, fog computing, | A multi-level DT-based system is proposed for the real-time monitoring, decision-making and control of a production logistics |

| | | monitor and evaluate the real-time operation status | edge computing | synchronisation system. A cloud-fog-edge-based DT control framework is developed to address the large computation scale, low computation efficiency and high information delay issues at the decision level. |
|---|---|---|---|---|
| [64] | Logistics | A systematic literature review of blockchain applications in supply chains, logistics and transport management | Blockchain | The four clusters of the blockchain (technology, trust, trade, and traceability/transparency) are discussed for the applications of the supply chain, logistics, and transport, such as increased synchronisation across logistics entities and enhanced trust by sharing secure and customised transport data. |
| [65] | Logistics | The lack of standard methodology for the design of a strategy to develop and validate a blockchain solution | Blockchain | A standard GUEST methodology is proposed to design real-world fresh food logistics using a case-utilising blockchain solution for reducing logistics costs and optimising operations. |

*3.2.4 Blockchain-based Digital Twin Warehouse and Logistics.* Warehouses have evolved rapidly in the last few decades, from the traditional warehouses that primarily utilise a manual handling system to the semi-automated warehouses using partially automated storage solutions and a Warehouse Management System (WMS) to today's automated warehouses integrating a number of Industry 4.0 technologies.

However, DTs can be implemented to optimise warehouse operations and management while minimising disruptions to the working tasks. The digital twin warehouse (DTW) is a bidirectional simulation model of the physical structure of the warehouse, which will be updated continuously with new data collected from smart sensors and RFID systems deployed around the warehouse. According to McKinsey, the global budget for the companies on warehousing a year is around 350 billion dollars. With the implementation of DTW, warehouse efficiency is predicted to increase by 20–25%, helping companies significantly save their operating and capital expenses [79]. Leveraging blockchain technology into DTW can further improve the security and efficiency of the WMS, especially concerning data management.

Blockchain allows warehouses, distribution centres, suppliers, manufacturers and other supply chain parties to interact with each other through permanent records of all transactions, which are stored and tracked within the decentralised blockchain network. Besides, the DTW can utilise smart contracts to automate verifications and payment. Thus, the blockchain-based DTW can significantly improve warehouse prediction and management accuracy and visibility.

A warehouse is built where goods or raw materials are stored for production or sale. Meanwhile, logistics refersto the inbound and outbound flow of goods transported in and out of a warehouse. Therefore, the traceability of data, such as location information, logistics networks, and the state of goods, is one of the most important concerns during transportation or internal movement within the warehouse. DTs can be created for the physical logistics network to achieve this goal, and the digital replica of physical assets can provide insights about the status, potential risks, and precise positionings during shipment and transportation. Besides, shippingcontainers connected to the blockchain and embedded with IoT sensors can bring complete transparency tothe logistics process. The items can be tracked and traced with the help of IoT sensors through creatinga DT on the blockchain platform. This ensures each authorised supply chain party involved in the logistics process has complete information regarding the items, containers, and conveyances. Table 7 discusses the main technologies applied and key findings of BS-DSCT research related to blockchain-based DT warehouses and logistics.

On the otherhand, the documents needed for the transportation process and the DT data collected from the IoT sensors deployed in each item can be transmitted and stored on the blockchain distributed ledgers to eliminate the risks of forgery or duplication. Then, data, documents and transactions can be tracked digitally and immutably [80]. Figure 9 shows a scenario of shipping goods, where several parties are involved, including manufacturers, suppliers, shippers, third-party logistics, warehouses, carriers and consignees. Eachequipment or system related to each party is embedded with multiple IoT sensors and has a digital replica. Then, DT data are transmitted to and stored in the blockchain network for tracking and validating. Besides, each transaction and document that occurred during the logistics process will generate apermanent ledger record that is easily validated by anyone authorised to access the blockchain network. Then, the network members can establish a transparent and efficient platform for managing all documents and transactions involved in the logistics process.

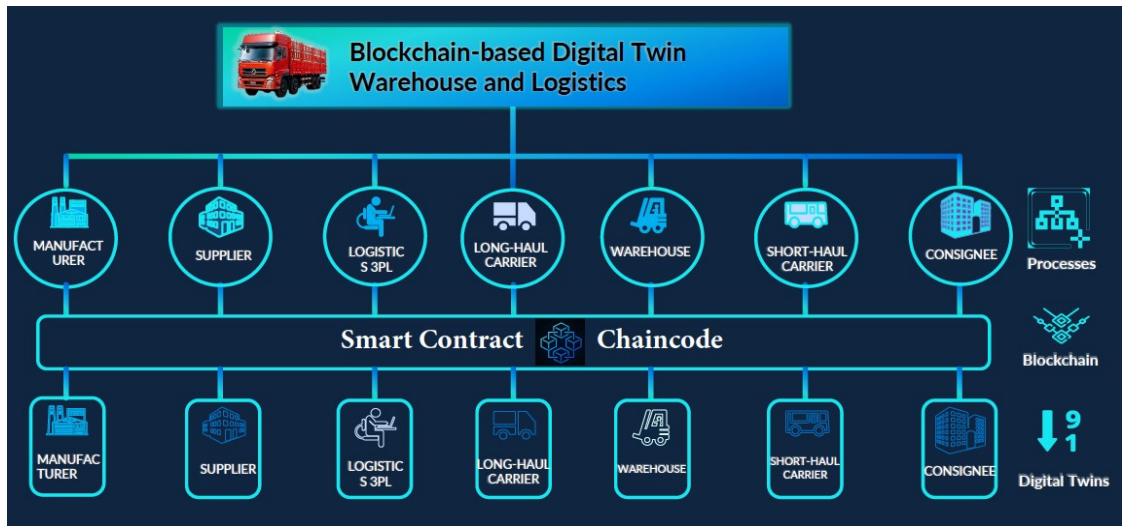

Fig. 9. Blockchain-based DT warehouse and logistics

## 4  FUTURE RESEARCH OPPORTUNITIES

Based on the literature review, we put forward future research opportunities.

- **Establishment of a BC-DSCT Platform**. The existing studies, such as [30] and [27], have developed DApps to achieve the decentralised data sharing and management of DT components and their associated data. By leveraging their work, future research can develop a BC-DSCT platform or DApp, which enables complete solutions of decentralised management, storage, and sharing of DT data while transmitting among the physical and digital spaces and then uploading storing on the blockchain network. The transactions of DT data among supply chain parties can also be tracked and verified in a transparent and tamper-proof environment. Considering that most participants within the supply chain system can constitute a consortium, implementing a consortium blockchain is the most suitable for the development of BC-DSCT DApps. Thus, we suggest utilising Hyperledger Fabric to establish the platform to realise the aforementioned functions and perform better scalability in a permissioned blockchain. Besides, it is necessary to develop feasible approaches that combine on-chain and off-chain storage to deal with a huge amount of DT data, significantly reducing latency and computing power compared to uploading all data onto the blockchain network.
- **Development of Data Management Methods**. The increasing DT data stored in the blockchain network will reduce the query efficiency of the blockchain, and the latency of query might cause security issues that need to be improved in future work regarding data management of DTs on blockchain systems [27]. Suggested future directions focus on several aspects. First, current data models applied in various blockchain platforms are not suitable for complex supply chain applications. Hence, we need to create a unified model to support different types of DT data storage in a single blockchain platform to protect data privacy. Second, each blockchain platform used by supply chain organisations will have different data schemas, which might cause data authenticity issues. Hence, it requires the development of a cross-chain network utilising smart contracts run across various blockchain systems. Third, a great amount of DT data generated from the interaction of the physical and digital spaces in SCM operation stored on the blockchain network is infeasible due to several limitations, including the storage size, transaction latency, and throughput. Therefore, it is important to develop a novel blockchain system integrated with a built-in offline storage method, which can deal with a huge amount of DT data [81].

- **Design of Reference Architecture.** A reference architecture provides the best practice of IT solutions by utilising an easy-to-understand format. However, current research lacks recommended structures for the integration of blockchain and DTs. Thus, it is necessary to establish a reference architecture for guiding the implementation of BC-DSCT to effectively address the interoperability issue. This should define a series of aspects, such as the level of integration, supply chain application, and data pattern, for the development, operation, simulation, and prediction of BC-DSCT solutions. The reference architecture should answer questions regarding how different levels of integration of blockchain and DTs can address specific issues in SCM, where and by what means to utilise blockchain and DTs in various parties and processes within the supply chain industry, and what data pattern could be used to synchronise and transmit DT data storing on the blockchain network. Similar research was conducted by Zuo [82], who constructed a reference architecture to integrate blockchain to industry 4.0 smart manufacturing. Future research can focus on establishing a more comprehensive reference architecture for BC-DSCT by adding more aspects to achieve the best practice in supply chains.

- **Implementation of BC-DSCT Standards**. Standardisation refers to the process of technical standards and consensus development and implementation for different supply chain parties to develop a unified system through all components of a product or system created and measured uniformly. However, supply chain standardisation involves building standards among supply chain organisations worldwide, which requires changes or restructuring of business processes. On the other hand, ISO [83] has released a number of international blockchain standards from 2019, including interactions between smart contracts and distributed ledger technology systems, privacy and personally identifiable information protection considerations, the security management of digital asset custodians, and blockchain vocabulary. Meanwhile, ISO and IEC [84] jointly involved themselves in the standardisation of DTs, including the term definition, system architecture, and reference model of DTs. However, the lack of industry standards for adopting blockchain and DTs in SCM will cause inefficiency and data inconsistency issues. Thus, it is necessary to develop relevant standards for BC-DSCT regarding technical guidelines, adoption requirements, evaluation criteria, instances, and capabilities. For example, seamless integration of DT data is required by implementing an open data structure utilising a uniform linked data language and several interoperate, agile, and scalable blockchains to improve interoperability and efficiency of BC-DSCT [36, 85]. Future researchers can also develop unified data and model standards for designing, developing, and implementing blockchain and DTs integration platforms suitable for supply chain applications.

- **Integration of BC-DSCT with Other Technologies**. Except for the perfect match between blockchains and DTs, integrating several other technologies can also significantly accelerate the adoption in SCM. Current research, such as [86-88], has integrated blockchains and DTs with federated learning to minimise costs and product delivery times while [36] discussing the combination of blockchain-based DTs and AI. IoT technology is closely associated with BC-DSCT since IoT sensors are widely deployed in each product within the supply chains to send data to the digital counterpart and receive data for maintenance. Meanwhile, IoT devices can utilise blockchains to achieve autonomous transactions and improve security and privacy. Therefore, future research can focus on conducting deeper research on the integration of BC-DSCT with IoT, federated learning, edge computing and AI. Besides, leveraging BC-DSCT and big data for predictive manufacturing is a promising direction in the future [89].

## 5 CONCLUSION

Blockchain and DTs are perfectly compatible in the domain of SCM, especially for the implementation of smart manufacturing, intelligent maintenance, blockchain-based DT shop floors, warehouses and logistics, due to the characteristics of improved security and efficiency of data management. In this paper, we conducted a comprehensive literature review of the blockchain-based DT integration for SCM. Based on the literature review, we put forward future research directions to guide the future improvement of BC-DSCT. We believe that this literature survey addresses the gap of the integration of blockchain and DTs with SCM and contributes to the development of DT adoption by leveraging blockchain technology to optimise the data management process.

## REFERENCES

[1] Abid Khan, Furqan Shahid, Carsten Maple, Awais Ahmad, and Gwanggil Jeon. Towards smart manufacturing using spiral digitaltwin framework and twinchain. *IEEE Transactions on Industrial Informatics*, 18(2), 1359-1366, 2020

[2] Gartner. Gartner survey reveals digital twins are entering mainstream use. https://www.gartner.com/en/newsroom/press-releases/2019-02-20-gartner-survey-reveals-digital-twins-are-entering-mai. Accessed November 24, 2021.

[3] Jiewu Leng, Dewen Wang, Weiming Shen, Xinyu Li, Qiang Liu, and Xin Chen. Digital twins-based smart manufacturing systemdesign in industry 4.0: A review. *Journal of Manufacturing Systems*, 60:119–137, 2021.

[4] Michael Grieves. Digital twin: manufacturing excellence through virtual factory replication. *White paper*, 1:1–7, 2014.

[5] Edward Glaessgen and David Stargel. The digital twin paradigm for future nasa and us air force vehicles. In *53rd AIAA/ASME/ASCE/AHS/ASC structures, structural dynamics and materials conference 20th AIAA/ASME/AHS adaptive structures conference 14th AIAA*, page 1818, 2012.

[6] Pethuru Raj. Empowering digital twins with blockchain. *Advances in Computers*, 121:267, 2021.

[7] Lei Hou, Shaoze Wu, Guomin Kevin Zhang, Yongtao Tan, and Xiangyu Wang. Literature review of digital twins applications inconstruction workforce safety. *Applied Sciences*, 11(1):339, 2021.

[8] Greyce N Schroeder, Charles Steinmetz, Carlos E Pereira, and Danubia B Espindola. Digital twin data modeling with automationmland a communication methodology for data exchange. *IFAC-PapersOnLine*, 49(30):12–17, 2016.

[9] Kazi Masudul Alam and Abdulmotaleb El Saddik. C2ps: A digital twin architecture reference model for the cloud-based cyber-physicalsystems. *IEEE access*, 5:2050–2062, 2017.

[10] Haya R Hasan, Khaled Salah, Raja Jayaraman, Mohammed Omar, Ibrar Yaqoob, Saša Pesic, Todd Taylor, and Dragan Boscovic. Ablockchain-based approach for the creation of digital twins. *IEEE Access*, 8:34113–34126, 2020.

[11] Barbara Rita Barricelli, Elena Casiraghi, and Daniela Fogli. A survey on digital twin: definitions, characteristics, applications, anddesign implications. *IEEE access*, 7:167653–167671, 2019.

[12] Werner Kritzinger, Matthias Karner, Georg Traar, Jan Henjes, and Wilfried Sihn. Digital twin in manufacturing: A categoricalliterature review and classification. *IFAC-PapersOnLine*, 51(11):1016–1022, 2018.

[13] Stefan Boschert and Roland Rosen. Digital twin—the simulation aspect. In *Mechatronic futures*, pages 59–74. Springer, 2016.

[14] Elisa Negri, Luca Fumagalli, and Marco Macchi. A review of the roles of digital twin in

cps-based production systems. *Procedia Manufacturing*, 11:939–948, 2017.

[15] Satoshi Nakamoto. Bitcoin: A peer-to-peer electronic cash system. *Decentralized Business Review*, page 21260, 2008.

[16] Abhishek Srivastava, Pronaya Bhattacharya, Arunendra Singh, Atul Mathur, U Pradesh, and U Pradesh. A systematic review on evolution of blockchain generations. *International Journal of Information Technology and Electrical Engineering*, 7(6):1–8, 2018.

[17] Arthur Carvalho. A permissioned blockchain-based implementation of lmsr prediction markets. *Decision Support Systems*, 130:113228, 2020.

[18] Arthur Carvalho. Bringing transparency and trustworthiness to loot boxes with blockchain and smart contracts. *Decision Support Systems*, 144:113508, 2021.

[19] Aaron Wright and Primavera De Filippi. *Blockchain and the law: the rule of code*. Harvard University Press, 2018.

[20] Florian Glaser, Florian Hawlitschek, and Benedikt Notheisen. Blockchain as a platform. In *Business transformation through Blockchain*, pages 121–143. Springer, 2019.

[21] Tien Tuan Anh Dinh, Rui Liu, Meihui Zhang, Gang Chen, Beng Chin Ooi, and Ji Wang. Untangling blockchain: A data processing view of blockchain systems. *IEEE transactions on knowledge and data engineering*, 30(7):1366–1385, 2018.

[22] Massimo Di Pierro. What is the blockchain? *Computing in Science & Engineering*, 19(5):92–95, 2017.

[23] Valentina Gatteschi, Fabrizio Lamberti, Claudio Demartini, Chiara Pranteda, and Victor Santamaria. To blockchain or not to blockchain: That is the question. *IT Professional*, 20(2):62–74, 2018.

[24] Shuchih E Chang and Yichian Chen. When blockchain meets supply chain: A systematic literature review on current development and potential applications. *IEEE Access*, 8:62478–62494, 2020.

[25] Pingcheng Ruan, Gang Chen, Tien Tuan Anh Dinh, Qian Lin, Beng Chin Ooi, and Meihui Zhang. Fine-grained, secure and efficient data provenance on blockchain systems. *Proceedings of the VLDB Endowment*, 12(9):975–988, 2019.

[26] Konstantinos Christidis and Michael Devetsikiotis. Blockchains and smart contracts for the internet of things. *Ieee Access*, 4:2292–2303, 2016.

[27] Sihan Huang, Guoxin Wang, Yan Yan, and Xiongbing Fang. Blockchain-based data management for digital twin of product. *Journal of Manufacturing Systems*, 54:361–371, 2020.

[28] Fei Tao, Yongping Zhang, Ying Cheng, Jiawei Ren, Dongxu Wang, Qinglin Qi, and Pei Li. Digital twin and blockchain enhanced smart manufacturing service collaboration and management. *Journal of Manufacturing Systems*, 2020.

[29] Sabah Suhail, Rasheed Hussain, Raja Jurdak, and Choong Seon Hong. Trustworthy digital twins in the industrial internet of things with blockchain. *IEEE Internet Computing*,

2021.

[30] Benedikt Putz, Marietheres Dietz, Philip Empl, and Günther Pernul. Ethertwin: Blockchain-based secure digital twin information management. *Information Processing & Management*, 58(1):102425, 2021.

[31] Marietheres Dietz, Benedikt Putz, and Günther Pernul. A distributed ledger approach to digital twin secure data sharing. In *IFIP Annual Conference on Data and Applications Security and Privacy*, pages 281–300. Springer, 2019.

[32] IBM. Blockchain for supply chain solutions. https://www.ibm.com/blockchain/supply-chain. Accessed November 17, 2021.

[33] Tim Payne. Digital planning requires a digital supply chain twin. https://www.gartner.com/en/doc/375403-supply-chain-brief-digital-planning-requires-a-digital-supply-chain-twin. Accessed December 2, 2021.

[34] Shan Ren, Xibin Zhao, Binbin Huang, Zhe Wang, and Xiaoyu Song. A framework for shopfloor material delivery based on real-time manufacturing big data. *Journal of Ambient Intelligence and Humanized Computing*, 10(3):1093–1108, 2019.

[35] Benoit Lheureux, W. Roy Schulte, and Alfonso Velosa. Why and how to design digital twins. https://www.gartner.com/en/documents/3888980/why-and-how-to-design-digital-twins. Accessed December 2, 2021.

[36] Sabah Suhail, Rasheed Hussain, Raja Jurdak, Alma Oracevic, Khaled Salah, Raimundas Matulevičius, and Choong Seon Hong. Blockchain-based digital twins: Research trends, issues, and future challenges. *ACM Computing Surveys (CSUR)*, 2021.

[37] Ibrar Yaqoob, Khaled Salah, Mueen Uddin, Raja Jayaraman, Mohammed Omar, and Muhammad Imran. Blockchain for digital twins: Recent advances and future research challenges. *IEEE Network*, 34(5):290–298, 2020.

[38] Piotr F Borowski. Digitization, digital twins, blockchain, and industry 4.0 as elements of management process in enterprises in the energy sector. *Energies*, 14(7):1885, 2021.

[39] StartUs Insights. 4 top digital twin startups impacting the energy industry. https://www.startus-insights.com/innovators-guide/4-top-digital-twin-startups-impacting-energy-industry/. Accessed November 20, 2021.

[40] Emergen Research. Top 10 digital twin companies impacting industry 4.0 innovations in 2021. https://www.emergenresearch.com/blog/top-10-digital-twin-companies-impacting-industry-4-0-innovations-in-2021. Accessed November 20, 2021.

[41] Cankal Altun, Bulent Tavli, and Halim Yanikomeroglu. Liberalization of digital twins of iot-enabled home appliances via blockchains and absolute ownership rights. *IEEE Communications Magazine*, 57(12):65–71, 2019.

[42] Khaled Salah, M Habib Ur Rehman, Nishara Nizamuddin, and Ala Al-Fuqaha. Blockchain for ai: Review and open research challenges. *IEEE Access*, 7:10127–10149, 2019.

[43] Jiewu Leng, Douxi Yan, Qiang Liu, Kailin Xu, J Leon Zhao, Rui Shi, Lijun Wei, Ding

Zhang, and Xin Chen. Manuchain: Combining permissioned blockchain with a holistic optimization model as bi-level intelligence for smart manufacturing. *IEEE Transactions on Systems, Man, and Cybernetics: Systems*, 50(1):182–192, 2019.

[44] Chao Zhang, Guanghui Zhou, Han Li, and Yan Cao. Manufacturing blockchain of things for the configuration of a data-and knowledge-driven digital twin manufacturing cell. *IEEE Internet of Things Journal*, 7(12):11884–11894, 2020.

[45] Claudio Mandolla, Antonio Messeni Petruzzelli, Gianluca Percoco, and Andrea Urbinati. Building a digital twin for additive manufacturing through the exploitation of blockchain: A case analysis of the aircraft industry. *Computers in Industry*, 109:134–152, 2019.

[46] Chiara Cimino, Elisa Negri, and Luca Fumagalli. Review of digital twin applications in manufacturing. *Computers in Industry*, 113:103130, 2019.

[47] Qinglin Qi and Fei Tao. Digital twin and big data towards smart manufacturing and industry 4.0: 360 degree comparison. *Ieee Access*, 6:3585–3593, 2018.

[48] Jiewu Leng, Shide Ye, Man Zhou, J Leon Zhao, Qiang Liu, Wei Guo, Wei Cao, and Leijie Fu. Blockchain-secured smart manufacturing in industry 4.0: A survey. *IEEE Transactions on Systems, Man, and Cybernetics: Systems*, 51(1):237–252, 2020.

[49] Weichao Luo, Tianliang Hu, Yingxin Ye, Chengrui Zhang, and Yongli Wei. A hybrid predictive maintenance approach for cnc machine tool driven by digital twin. *Robotics and Computer-Integrated Manufacturing*, 65:101974, 2020.

[50] Qiuan Chen, Zhenwei Zhu, Shubin Si, and Zhiqiang Cai. Intelligent maintenance of complex equipment based on blockchain and digital twin technologies. In *2020 IEEE International Conference on Industrial Engineering and Engineering Management (IEEM)*, pages 908–912. IEEE, 2020.

[51] Jay Lee, Jun Ni, Jaskaran Singh, Baoyang Jiang, Moslem Azamfar, and Jianshe Feng. Intelligent maintenance systems and predictive manufacturing. *Journal of Manufacturing Science and Engineering*, 142(11):110805, 2020.

[52] Itxaro Errandonea, Sergio Beltrán, and Saioa Arrizabalaga. Digital twin for maintenance: A literature review. *Computers in Industry*, 123:103316, 2020.

[53] Cunbo Zhuang, Tian Miao, Jianhua Liu, and Hui Xiong. The connotation of digital twin, and the construction and application method of shop-floor digital twin. *Robotics and Computer-Integrated Manufacturing*, 68:102075, 2021.

[54] Zhuang Cunbo, Jianhua Liu, and Hui Xiong. Digital twin-based smart production management and control framework for the complex product assembly shop-floor. *The international journal of advanced manufacturing technology*, 96(1-4):1149–1163, 2018.

[55] Fei Tao and Meng Zhang. Digital twin shop-floor: a new shop-floor paradigm towards smart manufacturing. *Ieee Access*, 5:20418–20427, 2017.

[56] Pedro Daniel Urbina Coronado, Roby Lynn, Wafa Louhichi, Mahmoud Parto, Ethan

Wescoat, and Thomas Kurfess. Part data integration in the shop floor digital twin: Mobile and cloud technologies to enable a manufacturing execution system. *Journal of manufacturing systems*, 48:25–33, 2018.

[57] Dominik Welte, Axel Sikora, Daniel Schönle, Jan Stodt, and Christoph Reich. Blockchain at the shop floor for maintenance. In *2020 International Conference on Cyber-Enabled Distributed Computing and Knowledge Discovery (CyberC)*, pages 15–22. IEEE, 2020.

[58] Kyu Tae Park, Yoo Ho Son, and Sang Do Noh. The architectural framework of a cyber physical logistics system for digital-twin-based supply chain control. *International Journal of Production Research*, 59(19):5721–5742, 2021.

[59] Fábio Coelho, Susana Relvas, and AP Barbosa-Póvoa. Simulation-based decision support tool for in-house logistics: the basis for a digital twin. *Computers & Industrial Engineering*, 153:107094, 2021.

[60] YH Pan, T Qu, NQ Wu, M Khalgui, and GQ Huang. Digital twin based real-time production logistics synchronization system in a multi-level computing architecture. *Journal of Manufacturing Systems*, 58:246–260, 2021.

[61] Jiewu Leng, Douxi Yan, Qiang Liu, Hao Zhang, Gege Zhao, Lijun Wei, Ding Zhang, Ailin Yu, and Xin Chen. Digital twin-driven joint optimisation of packing and storage assignment in large-scale automated high-rise warehouse product-service system. *International Journal of Computer Integrated Manufacturing*, 34(7-8):783–800, 2021.

[62] Guido Perboli, Stefano Musso, and Mariangela Rosano. Blockchain in logistics and supply chain: A lean approach for designing real-world use cases. *Ieee Access*, 6:62018–62028, 2018.

[63] Rudra Prasad Tripathy, Manoj Ranjan Mishra, and Satya Ranjan Dash. Next generation warehouse through disruptive iot blockchain. In *2020 International Conference on Computer Science, Engineering and Applications (ICCSEA)*, pages 1–6. IEEE, 2020.

[64] Mehrdokht Pournader, Yangyan Shi, Stefan Seuring, and SC Lenny Koh. Blockchain applications in supply chains, transport and logistics: a systematic review of the literature. *International Journal of Production Research*, 58(7):2063–2081, 2020.

[65] Shazhou Chen, Wei Meng, Weiyuan Xu, Zhuoqiang Liu, Jiachuang Liu, and Fengyan Wu. A warehouse management system with uav based on digital twin and 5g technologies. In *2020 7th International Conference on Information, Cybernetics, and Computational Social Systems (ICCSS)*, pages 864–869. IEEE, 2020.

[66] Pethuru Raj and Chellammal Surianarayanan. Digital twin: the industry use cases. In *Advances in Computers*, volume 117, pages 285–320. Elsevier, 2020.

[67] GE Digital. Digital twin-apply advanced analytics and machine learning to reduce operational costs and risks. https://www.ge.com/ digital/applications/digital-twin. Accessed November 20, 2021.

[68] Jose Antonio Marmolejo-Saucedo. Design and development of digital twins: A case study

in supply chains. *Mobile Networks and Applications*, 25:2141–2160, 2020.

[69] Preetha Evangeline et al. Digital twin technology for "smart manufacturing". In *Advances in Computers*, volume 117, pages 35–49. Elsevier, 2020.

[70] Nader Mohamed and Jameela Al-Jaroodi. Applying blockchain in industry 4.0 applications. In *2019 IEEE 9th annual computing and communication workshop and conference (CCWC)*, pages 0852–0858. IEEE, 2019.

[71] Logan D Sturm, Christopher B Williams, Jamie A Camelio, Jules White, and Robert Parker. Cyber-physical vulnerabilities in additive manufacturing systems: A case study attack on the. stl file with human subjects. *Journal of Manufacturing Systems*, 44:154–164, 2017.

[72] Nallapaneni Manoj Kumar and Pradeep Kumar Mallick. Blockchain technology for security issues and challenges in iot. *Procedia Computer Science*, 132:1815–1823, 2018.

[73] Ghosh Debabrata and Tan Albert. A framework for implementing blockchain technologies to improve supply chain performance. 2018.

[74] Dimitris Mourtzis and Michalis Doukas. Decentralized manufacturing systems review: challenges and outlook. In *Robust Manufacturing Control*, pages 355–369. Springer, 2013.

[75] Weiming Shen. Distributed manufacturing scheduling using intelligent agents. *IEEE intelligent systems*, 17(1):88–94, 2002.

[76] Simon Kranzer, Dorian Prill, Davood Aghajanpour, Robert Merz, Rafaela Strasser, Reinhard Mayr, Helmut Zoerrer, Matthias Plasch, and Robert Steringer. An intelligent maintenance planning framework prototype for production systems. In *2017 IEEE International Conference on Industrial Technology (ICIT)*, pages 1124–1129. IEEE, 2017.

[77] Ansys. Digital twin-transform your operations with data-driven and simulation-based digital twin software. https://www.ansys.com/ products/digital-twin. Accessed November 22, 2021.

[78] Ansys. Ansys collaborates with ge to drive digital twin value and deliver the promise of the industrial internet of things. https://www.prnewswire.com/news-releases/ansys-collaborates-with-ge-to-drive-digital-twin-value-and-deliver-the-promise-of-the-industrial-internet-of-things-300363787.html. Accessed November 22, 2021.

[79] Felipe Bustamante, Ashutosh Dekhne, Jörn Herrmann, and Vedang Singh. Improving warehouse operations—digitally. https://www.mckinsey.com/business-functions/operations/our-insights/improving-warehouse-operations-digitally. Accessed November 23, 2021.

[80] Naveen Joshi. The curious case of digital twins and blockchain. https://www.allerin.com/blog/the-curious-case-of-digital-twins-and-blockchain. Accessed November 24, 2021.

[81] Hoang Tam Vo, Ashish Kundu, and Mukesh K Mohania. Research directions in blockchain data management and analytics. In *EDBT*, pages 445–448, 2018.

[82] Zuo, Y. (2021). Making smart manufacturing smarter–a survey on blockchain technology

Blockchain-based Digital Twin for Supply Chain Management

in Industry 4.0. *Enterprise Information Systems*, *15*(10), 1323-1353.

[83] ISO. Blockchain standards. https://www.iso.org/search.html?q=blockchain&hPP=10&idx=all_en&p=0&hFR%5Bcategory%5D%5B0%5D=standard. Accessed November 26, 2021.

[84] IEC. Moving ahead on standardization for digital twin. https://www.iec.ch/blog/moving-ahead-standardization-digital-twin. Accessed November 26, 2021.

[85] Subin Shen. Blockchain and its standardization work. https://www.itu.int/en/ITU-T/academia/kaleidoscope/2017/Documents/special/SS02.pdf. Accessed December 10, 2021.

[86] Yunlong Lu, Xiaohong Huang, Ke Zhang, Sabita Maharjan, and Yan Zhang. Communication-efficient federated learning and permissioned blockchain for digital twin edge networks. *IEEE Internet of Things Journal*, 8(4):2276–2288, 2020.

[87] Yunlong Lu, Xiaohong Huang, Ke Zhang, Sabita Maharjan, and Yan Zhang. Low-latency federated learning and blockchain for edge association in digital twin empowered 6g networks. *IEEE Transactions on Industrial Informatics*, 17(7):5098–5107, 2020

[88] Li Jiang, Hao Zheng, Hui Tian, Shengli Xie, and Yan Zhang. Cooperative federated learning and model update verification in blockchain empowered digital twin edge networks. *IEEE Internet of Things Journal*, 2021

[89] Bao, J., Guo, D., Li, J., & Zhang, J. (2019). The modelling and operations for the digital twin in the context of manufacturing. *Enterprise Information Systems*, *13*(4), 534-556.